\documentclass{nle}
\usepackage{inputenc}
\usepackage{graphicx}
\usepackage{hyperref}
\usepackage{algorithm}
\usepackage{algpseudocode}

\title[Court Judgements Labeling Using Topic Modeling and Syntactic Parsing]
      {Court Judgements Labeling Using Topic Modeling and Syntactic Parsing}
\author[Yuchen Liu]
       {Yuchen Liu}

\pagerange{\pageref{firstpage}--\pageref{lastpage}}
\pubyear{2022}

\begin{document}

\label{firstpage}
\maketitle

\begin{abstract}
In regions that practice common law, relevant historical cases are essential references for sentencing. To help legal practitioners find previous judgement easier, this paper aims to label each court judgement by some tags. These tags are legally important to summarize the judgement and can guide the user to similar judgements. We introduce a heuristic system to solve the problem, which starts from Aspect-driven Topic Modeling and uses Dependency Parsing and Constituency Parsing for phrase generation. We also construct a legal term tree for Hong Kong and implemented a sentence simplification module to support the system. Finally, we propose a similar document recommendation algorithm based on the generated tags. It enables users to find similar documents based on a few selected aspects rather than the whole passage. Experiment results show that this system is the best approach for this specific task. It is better than simple term extraction method in terms of summarizing the document, and the recommendation algorithm is more effective than full-text comparison approaches. We believe that the system has huge potential in law as well as in other areas.
\end{abstract}

\section{Introduction}

\subsection{Problem Statement}

In regions that practice common law, the tribunal relies highly on previous court decisions to resolve disputes. Legal practitioners frequently need to find relevant legal judgements at work. Indexing each court judgements by labels can considerably increase the speed of court judgement recommendation, and can help users find other judgements with similar topics faster. 

This paper will focus on labeling every court judgement by some tags. There are existing paid websites that provide such tags, for example Westlaw Asia. However, they maintained the tags by recruiting legal experts to label manually, which is expensive and time-consuming. Plus, users cannot select the tags to find similar documents. A better solution must be developing an NLP system that can automatically extract legally important tags from a court judgement. It leads to the question that how to build such a system and how to find similar documents using these tags.

\subsection{Literature Review}

\begin{figure}[htp]
    \centering
    \includegraphics[width=\textwidth]{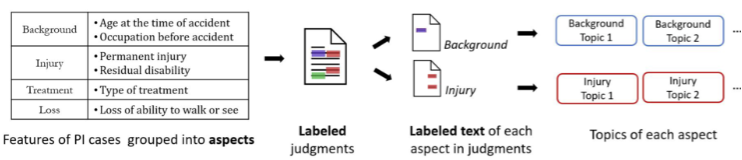}
    \caption{Aspect-driven Topic Modeling}
    \label{fig:2}
\end{figure}

Previous work of Wu, et al. \shortcite{wu22} focuses on semantic search and summarization using \textbf{Topic Modeling}. It proved that the Topic Modeling based on labeled aspects (background, injury, losses, and compensations for Personal Injury cases) shows the best summarization ability. Figure \ref{fig:2} shows the general process of Aspect-driven Topic Modeling. 

For each aspect of documents in the training set, a few significant topics will be generated. Based on the resultant topics, the similarity between paragraphs and aspects can be calculated using the normalized inner product of the word2vec  embeddings of the paragraphs and topics \cite{3}. This way, paragraphs that are related to each aspect can be found. By simply changing paragraphs to sentences in this algorithm, we can find the most relevant sentences to each aspect. These aspect sentences should comprehensively summarize the court judgement and will serve as the starting point of this project.

Two of the classical Syntactic Parsing approaches will be used in our model.

\textbf{Dependency Parsing} finds the asymmetric dependency relations between words. These relations can be represented as arrows pointing from the heads to the dependents. Usually, the heads are more grammatically important than dependents. For example, adjectives normally depend on nouns. The result should form a non-recurrent tree. We can backtrack from one leaf of the tree and find all its parent words.

\textbf{Constituency Parsing} uses phrase structure grammar to organize words into nested constituents. Each constituent is labeled by one of the predetermined grammar units, called POS Tags, for example, VB denotes verb phrase and PP denotes prepositional phrase.

\subsection{Objective}

The tags given to each judgement should fulfill the below properties.
\begin{enumerate}
    \item The tags should be/contain legally important terms.
    \item The tags are free of grammatical/content mistakes that will affect understanding.
    \item The tags can navigate the user to other similar documents.
    \item The tags can describe important aspects of the current judgement.
\end{enumerate}
As illustrated in 1.3, if we generate tags based on the aspect sentences, requirement 4 will be satisfied \cite{wu22}. Therefore, we will only focus on solving requirements 1-3 in the following methodology. Because of the regional difference of constitution, we will use Hong Kong as the example and use the \href{https://www.hklii.hk/eng/}{HKLII} website as the data source for historical judgements.

\subsection{Report Outline}

Chapter 2 will introduce two supporting modules and the main algorithm. The similar document recommendation algorithm will then be illustrated. Chapter 3 will show the experiment process and discuss the results. Limitations and the future work for the project will be pointed out as well. Finally, the conclusion is given in Chapter 4.

\section{Methodology}

This section will start by introducing the preparation work for the system, including constructing the legal concept tree and developing the sentence simplification module. Then, it will compare contextual phrases against legal terms to discuss if phrase generation algorithms are needed. At last, it will present a novel model for extracting legally important phrases from sentences. 

\subsection{Preparation Work}

We have created two standalone modules that support and improve the main algorithm. Both of the modules have been open-sourced on \href{https://github.com/hkulyc/legalAI}{GitHub}.

\begin{figure}[htp]
    \centering
    \includegraphics{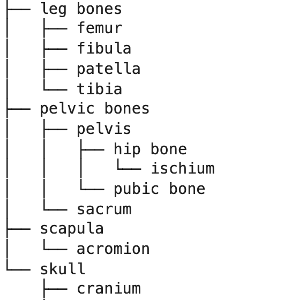}
    \caption{Part of the legal concepts tree}
    \label{fig:3}
\end{figure}

\textbf{Legal Concepts Tree} contains more than 13000 common legal terms used in Hong Kong Judiciary. It is adapted from the legal taxonomy developed by Sweet \& Maxwell. As shown in Figure \ref{fig:3}, it has a perfect tree structure where parent terms are more general than their children. The tree is optimized to store in a hashed map for faster key search. With this tree, we can use keyword matching to identify the legal terms in the sentence. If necessary, it is also useful for substituting the legal terms with more general ones.

\textbf{Sentence Simplification} is the task to split complex or compound sentences into simple sentences, each of which has only one pair of subject and verb. There are previous works on this task using constituency parsing, such as Garain et al. \shortcite{4} and Das et al. \shortcite{5}. We implement and improve the algorithm by Garain et al. \shortcite{4} to fit into the latest Stanford Constituency Parsing rule \cite{6,7}. This implementation is also intelligent to skip the proper names, such as "Fish and Chip" or "Johnson and Johnson". This module can help to simplify long sentences, which are common in court judgements, so that operations afterward can be more accurate.

\subsection{Tag Format Comparison}

One critical question to ask is that since the legal concept tree can help to find legal terms, is it necessary to take one step further to construct contextual phrases based on the legal terms? For instance, given the sentence "She suffered a fracture of the superior and inferior public rami of the right pelvis.", is tag "fracture" enough, or is tag "fracture of rami of pelvis" better? While phrase tags can contain more local contexts and information than word tags, they may sometimes be broken during generation. It is impetuous to draw any conclusion at this point, so one experiment is set up to test this question. The result is shown in chapter 3. The remaining sections in chapter 2 will focus on generating phrase tags only.

\subsection{Main Algorithm}

As stated in 1.4, Aspect-driven Topic Modelling will be used and the most relevant sentences to the aspects will be found first. The main algorithm solves the problem of extracting legally important phrases from each sentence.

\begin{algorithm}
\caption{The algorithm to transform a sentence into tags (variables are italic)}\label{alg:1}
\begin{algorithmic}
\Procedure{Generate Tags}{sentence $x$}
    \State $sentences$ $\gets$ split $x$ into simple ones, using split sentence module
    \State $tags \gets []$
    \State $N \gets n$
    \For{sentence $s$ in $sentences$}
        \State $dp\_tree$ $\gets$ the enhanced dependency tree of $s$ \Comment{A}
        \For{word $w$ in $s$}
            \If{$w$ in legal concepts tree}
                \State $words$ $\gets$ the smallest constituents containing $w$ \Comment{B}
                \For{$parent$ of $w$ in $dp\_tree$}
                    \State $words.append$($parent$, preposition betweeen $parent$ and $w$) \Comment{C}
                    \State $w \gets parent$
                \EndFor
                \State $tags.append$(join $words$ by " " in same order of sentence $x$)
            \EndIf
        \EndFor
    \EndFor
    \State \textbf{return} $tags$
\EndProcedure
\end{algorithmic}
\end{algorithm}

In step A, the Stanford Enhanced Dependency \cite{6} should be used. It differs from the Universal Dependency in containing prepositions inside relations. It will benefit the reconstruction of extracted words in step C. It should be noted that the relations starting with "acl:" should be eliminated as they will cause dependency cycles.
Step B finds the shortest subset of the sentence that contains the matching legal term. Based on Constituency Parsing, it is to find the shortest component with a tag that ends with "P", such as "NP", "VP". This will include the neighbor words of the legal term to make the result more specific. For example, given the constituency tree in Figure 2.3, assume that the word "condition" is found to be legal term. 

\begin{figure}[htp]
    \centering
    \includegraphics[width=\textwidth]{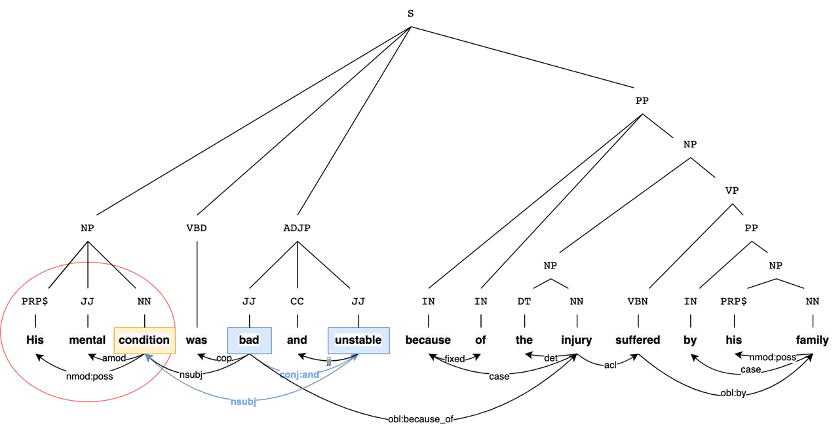}
    \caption{ Constituency Parsing Tree and Dependency Parsing Tree of the example sentence (arrows are pointing from parent to children)}
    \label{fig:4}
\end{figure}

Step B finds the smallest component "His mental condition". Without this step, the word "condition" itself might be too general and ambiguous.

Step C is to extract faraway related words with respect to the matching legal term using Dependency Parsing. Starting from the legal term w, the program will recursively find the parent of it in the dependency tree until the root is reached or a verb is met. Since the verb is connected to its subject using relation "obj", the condition therefore becomes meeting "obj". During the backtracking, the prepositions and conjunctions in the dependency relation should also be recorded. They are stored after ":" in the relation within the set {"nmod:", "obl:", "conj:", "advcl:"}. Finally, the words in the reconstructed phrase should have the same order as the original sentence. For the example of Figure 2.3, starting from the word "condition", "unstable" is found to be the first parent and "bad" is the second. There is a conjunction word "and" between the "bad" and "unstable", according to the relation "conj:and", so "and" should be inserted in between. As a result, the tag "His mental condition bad and unstable" is generated.

Overall, this algorithm is a hybrid approach that utilizes both Constituency Parsing in extracting neighboring phrase and Dependency Parsing in finding faraway relational words. To make the tags more fluent and natural to people, we also added several heuristic operations to the algorithm.
\begin{enumerate}
    \item If the last word of the recursive step C is a link verb, such as "is", "are", we should truncate it.
    \item If the smallest component found in step B does not help specify the matching term, for example only adding "the" in front of the matching term, step B should be skipped. In the implementation, the skipping condition is that the length of the component is two and the POS tag for the first word belongs to \{'DT', 'IN', 'PRP\$', 'CD'\}.
    \item Sometimes, the preposition "to" before a verb cannot be successfully recognized by the Dependency Parsing. We should add "to" in front of the word with POS Tag "VB".

\end{enumerate}

\subsection{Similar Documents Recommendation}
Assume we have obtained the tags, ultimately the user will choose a few of them to navigate to other judgements which may mention similar legal concepts. According to the user survey, users often have a few interested legal concepts in mind during searching. Therefore, this aspect-specific similar document matching may be more useful than comparing two judgements by the full text. To solve the problem, the simplest way is to directly invoke the default search engine and use the selected tags as search input. This way, the selected tags are compared with the whole passage of other documents. Another way is to compare the selected tags with tags from other documents. The theoretical basis of this approach is that the tags have summarized most aspects of a court judgement (Requirement 4 in 1.4). Narrowing the searching scope to tags only can help to eliminate interference from other unimportant texts, making the recommendation more accurate. The quality of this recommendation will be tested in the following experiment. Algorithm \ref{alg:2} shows the details of the recommendation algorithm by comparing selected tags with tags from other documents.

\begin{algorithm}[htp]
\caption{Recommendation Algorithm Details}\label{alg:2}
\begin{algorithmic}
\Function{CAL\_SIMILARITY}{$emb1$, $emb2$}
    \State \textbf{return} $emb1 \cdot emb2 / (\|emb1\| \cdot \|emb2\|)$
\EndFunction
\Procedure{Recommend Judgements}{selected tags $x$}
    \State $ret$ $\gets$ an empty max heap
    \ForAll{judgement $j$}
        \State $score \gets 0$ 
        \For{tag $t1$ of $x$}
            \State $e1 \gets$ embedding of $t1$
            \State $each\_score \gets 0$
                \For{tag $t2$ of $j$}
                    \State $e2 \gets$ embedding of $t2$
                    \State $each\_score \gets$ max($each\_score$, CAL\_SIMILARITY($e1$, $e2$))
                \EndFor
                \State $score$ += $each\_score$
        \EndFor
        \State add $j$ to $ret$ based on $score$
    \EndFor
    \State \textbf{return} $ret$
\EndProcedure
\end{algorithmic}
\end{algorithm}

Basically, the similarity between each selected tag and another document is equal to the maximum normalized inner product of all possible tag combinations. The similarity score between all selected tag and another document is the sum of such similarities.

\section{Evaluation}

This section detailly describes the experiment to evaluate the quality of the tags generated. Also, the experiment results will be revealed and the future work for the project will be pointed out.

\subsection{Experiment setup}

As mentioned in section 2.3, the main algorithm for generating phrase tags is a hybrid approach of Dependency Parsing and Constituency Parsing. To test whether the use of both parsing techniques is necessary, we set up a comparison experiment. For each court judgement, four tag sets are generated using 1. only Dependency Parsing 2. only Constituency Parsing 3. the Hybrid Approach 4. Word tags only. Each tag set starts from three starting sentences derived from Aspect-driven Topic Modeling. Each document along with all the resultant tags will be distributed to at least two participants to minimize personal bias. Participants are asked to select the tags that fulfill all three requirements in 1.4.
After getting the result from this experiment 1, we continue to test the tags format issue mentioned in section 2.2. The recommendation abilities of word tags and phrase tags are tested in this experiment 2. For each judgement, participants will be provided with the top 5 recommendations from word tags and phrase tags respectively in pairs. Participants are asked to find the better recommendation, which is more similar to the original judgement in each pair.

\subsection{Experiment Results}

We invited 7 participants to evaluate on tags of 15 independent court judgements. Table \ref{table:1} shows the results of experiment 1. The average number of tags generated by the four approaches per judgement and the percentage of good tags are provided. In total, 6.8*15 = 102 tags are generated by the hybrid method, and 1.76*15 = 26.3 of them are good.

\begin{table}[htp]
\begin{minipage}{\textwidth}
\begin{tabular}{lccc}
\hline\hline
Methods & \# of tags & \# of good tags & Good tags rate\\
\hline
Only Dependency Parsing & 8\hpt53 & 1\hpt74 & 0\hpt204\\
Only Constituency Parsing & 5\hpt80 & 1\hpt56 & 0\hpt268\\
Hybrid approach & 6\hpt80 & 1\hpt76 & 0\hpt258\\
Word tags & 10\hpt60 & 1\hpt58 & 0\hpt149\\
\hline\hline
\end{tabular}
\vspace{-1\baselineskip}
\end{minipage}
\caption{Statistics of good tags per judgement}
\label{table:1}
\end{table}

From the statistics, we can conclude that the hybrid approach generated the greatest number of good tags for each judgement (1.76). Also, although the constituency parsing approach generated the least good tags (1.56), the percentage of the good tags is the highest (0.268). Considering the purpose of the model, which is to generate tags for the users to choose, we should prefer the absolute number of good tags to the rate of it. Therefore, both Dependency Parsing and Constituency Parsing is necessary. The hybrid approach is the best phrase generation method. As for the word tags, the number of good tags is less than phrases, and the rate of good tags is the lowest. As a result, legal terms are not comparable to contextual legal phrases in terms of fulfilling the requirements of tags.

\begin{table}[htp]
\begin{minipage}{\textwidth}
\begin{tabular}{lcc}
\hline\hline
Cases & \# of pairs & Rate\\
\hline
Phrase tags better & 0\hpt88 & 0\hpt175\\
Word tags better & 1\hpt67 & 0\hpt336\\
Both are good & 0\hpt99 & 0\hpt198\\
Neither is similar & 1\hpt46 & 0\hpt291\\
Total & 5\hpt00 & 1\hpt000\\
\hline\hline
\end{tabular}
\vspace{-1\baselineskip}
\end{minipage}
\caption{Statistics of better recommendations per document}
\label{table:2}
\end{table}

Table \ref{table:2} shows the results of the recommendation ability of word tag and phrase tag in experiment 2. Since 5 pairs of external documents are recommended for each judgement, there are 15*5=75 pairs of comparison tested. Here, we use the hybrid approach for the phrase tag generation.

From the table, we can find that the frequency of word tags being better is higher than the number of phrase tags being better. This indicates that there is still room for improvement for the recommendation algorithm based on phrase tags. The possible direction will be discussed in section 3.4.

\subsection{Limitations}

The percentage of good tags from the hybrid approach is around 26\%. Although the purpose of the task does not require tags to have an extremely high acceptance rate, higher rate can definitely resolve some misunderstandings and improve user experience. Based on the feedbacks from experiment participants, the reason of bad tags mostly owes to not fulfilling requirement 2 from section 1.4. They often present broken or confusing meanings. The root of this problem is obvious. Both Dependency Parsing and Constituency Parsing are syntactic parsing, indicating that they do not consider semantic information enough. For example, the correct punctuation of sentence "Drive the lorry for loading and unloading goods" should be "Drive the lorry for loading goods" and "… for unloading goods". However, the constituency parsing splits like "Drive the lorry for loading" and "… for unloading goods". The parsing simply finds the result with the highest probability in grammar regardless of the obvious parallelism of antonyms: "loading" and "unloading". Therefore, to improve the main algorithm, essentially a new parsing method that maximizes both grammatical and semantic probability is needed.
Furthermore, since the main algorithm starts from most relevant sentences of each aspect, the result of AdTM will affect the final performance to large extent. Therefore, it is also necessary for the algorithm to find better document summarization methods alternative to AdTM that can elevate the upper limit.

\subsection{Future Work}

One promising way to improve the recommendation based on phrase tags is giving weights to different tag combinations. As illustrated in 2.4, we are summing over the maximum of similarities of combinations by considering all tags equal. However, different tags come from different sentences that belongs to topics in four different aspects. If two tags come from the same aspect or even the same topic, their similarity score should be assigned a higher weight. For example, two tags both describing the losses of the personal injury cases should be naturally closer than two that describes different aspects.
What's more, currently training sets and testing sets are all court judgements of Personal Injury cases. The effectiveness of this system on other case types is not tested. Future work should focus on applying to model to various types of cases.
Finding more applications for the system is also an interesting topic. This model is designed for the HKLII website. However, after a simple modification of the Legal Concept Tree in section 2.1 to fit into local context, this model can be applied to other law systems. Furthermore, other professional areas that have similar concept tree structures can easily adapt the model to their advantage.

\section{Conclusion}

This report introduced the model that labels court judgement. The system uses Aspect-driven Topic Modeling as the starting point. The main algorithm in the system that extracts phrase tags from a sentence is a hybrid method of Dependency Parsing and Constituency Parsing. Experiment results show that these tags outperform general legal terms. Although the aspect-specific recommendation is more useful than full passage comparison, it sometimes did not recommend the most similar documents compared with the benchmark algorithm. Future work should focus on the improvement of the recommendation algorithm for phrase tags, the seeking of alternative system components and the discovery of more use cases for the system.

\label{lastpage}

\begin{thebibliography}{}

  \bibitem[\protect\citename{Zhong {\it et al.}}2020]{zhong20}
   H. Zhong, C. Xiao, C. Tu, et al., "How does NLP benefit legal system: A summary of legal artificial intelligence," arXiv preprint arXiv, 2020.
  \bibitem[\protect\citename{Wu {\it et al.}}2022]{wu22}
   T. H. Wu, B. Kao, et al., "Semantic Search and Summarization of Judgments Using Topic Modeling," in Legal Knowledge and Information Systems, Erich Schweighofer, 2022, pp.100-106.
  \bibitem[\protect\citename{Mikolov {\it et al.}}2013]{3}
   T. Mikolov, I. Sutskever, K. Chen, G. Corrado, and J. Dean, "Distributed representations of words and phrases and their compositionality," Proceedings of the 26th International Conference on Neural Information Processing Systems - Volume 2 (NIPS'13), 2013, pp. 3111–3119.
  \bibitem[\protect\citename{Garain {\it et al.}}2019]{4}
   A. Garain, A. Basu, R. Dawn and S. K. Naskar, "Sentence Simplification using Syntactic Parse trees," 2019 4th International Conference on Information Systems and Computer Networks (ISCON), 2019, pp. 672-676.
   \bibitem[\protect\citename{Das {\it et al.}}2018]{5}
   B. Das, M. Majumder and S. Phadikar, "A Novel System for Generating Simple Sentences from Complex and Compound Sentences," International Journal of Modern Education and Computer Science, 2018, pp. 57.
   \bibitem[\protect\citename{Marneffe {\it et al.}}2021]{6}
   M. C. Marneffe, C. D. Manning, J. Nivre, D. Zeman, "Universal Dependencies," Computational Linguistics, 2021, pp. 255–308.
   \bibitem[\protect\citename{Schuster {\it et al.}}2016]{7}
   S. Schuster, C. D. Manning, "Enhanced English Universal Dependencies: An Improved Representation for Natural Language Understanding Tasks," Proceedings of the Tenth International Conference on Language Resources and Evaluation (LREC'16), 2016, pp. 2371-2378.

\end{thebibliography}
\end{document}